\documentstyle[12pt,emlines2]{article}
\newcommand{\tr}[1]{\,{\rm tr}\,#1\,}
\renewcommand{\theequation}{\thesection.\arabic{equation}}
\begin{document}
\title{
\begin{flushright}
{\small SMI-15-97 }
\end{flushright}
\vspace{2cm}
Knots and Matrix Models}
\author{
I.Ya.Aref'eva
\thanks
{ e-mail: arefeva@genesis.mi.ras.ru}
$~$ and $~~$ I.V.Volovich
\thanks{e-mail: volovich@genesis.mi.ras.ru}
$$~$$
$$~$$
\\Steklov Mathematical Institute\\
Gubkin St. 8, GSP-1, 117966, Moscow, Russia}
\date {$~$}
\maketitle
\begin {abstract}
We consider a matrix model with $d$ matrices $N\times N$
and show that in the limit $N\to\infty$
and $d\to 0$ the model describes the knot diagrams.
The same limit in matrix string theory is also discussed.
We speculate that a prototypical M(atrix)  without
matrix theory exists in void.
\end {abstract}

\newpage
\section{Introduction}
\setcounter{equation}{0}
Theory of knots \cite {Ati,Kau} is used in low dimensional topology and also
in physics, chemistry and biology.
Recently Faddeev and Niemi \cite {FN} have suggested
that in certain relativistic quantum field theories knotlike configurations
may appear as stable solitons. The
remarkable progress in the classification of knots
originated from Jones' and Vassiliev's invariants
was related with the application of von Neumann algebras, Yang-Baxter
equations and singularity theory, for a review see \cite {Bir,BGRT}.
Witten \cite {Wit} used methods of quantum field theory
in the theory of knots by considering  the Wilson
loops in the Chern-Simons gauge theory as knots.

In this note we describe  another application of quantum field theory to
the theory of knots.
We consider a matrix model with $d$ matrices $N\times N$
and show that in the limit $N\to\infty$
and $d\to 0$ the model describes the knot diagrams.

In the traditional approach to knot theory one deals with a single
knot. According to  Arnold's \cite {Arn} and Vassiliev's \cite {Vas}
point of view one has to consider not a single knot but a space of all
knots (cohomology of complements to discriminants). In this note
from the very beginning we don't have neither knots nor a three
dimensional space. We start from a matrix model in zero-dimensional
space and knots appear in the limit of large matrices.
The information about knots is encoded into the form of the Lagrangian
of the matrix model. In some sense this approach reminds the matrix
approach to superstring theory \cite {BFSS,Per,IKKT,DVV} where space-time
is represented as the moduli space of vacuum and strings appear
in the large $N$ limit.

\section{Matrix Model and Knots}
\setcounter{equation}{0}

Let be given positive integers $N$ and $d$ and let $
A_{\mu }
=(A_{\mu}^{ ij}) $ and $B_{\mu}=(B_{\mu}^{ ij}),~
i,j=1,...,N$ are $N\times N$ Hermitian matrices, $
A_{\mu}^{ *}=
A_{\mu},~ B_{\mu}^*=B_{\mu}$. Here $\mu=1,...,d$.
The matrix model has the following partition function
\begin {equation} 
                                                          \label {1.1}
Z=Z(N,d,g)=\int e^{iS}dA dB
\end   {equation} 
where the Lagrangian is
\begin {equation} 
                                                          \label {1.2}
S=Tr(A_{\mu}B_{\mu})+\frac{g}{2N}Tr(A_{\mu}
B_{\nu}
A_{\mu}B_{\nu})
\end   {equation} 
and we assume the summation over the repeating indices.
Here $g$ is a real parameter (coupling constant) and the measure
\begin {equation} 
dA=\prod_{\mu =1}^d
((\prod_{1\leq i\leq j\leq N} dReA_{\mu}^{ ij})\prod_{1\leq i
<j\leq N}dImA_{\mu}^{ ij})
\end   {equation} 
There exists a natural extension of the function $Z(N,d,g)$
from integers to real values of the parameter $d$. Actually
one has the following formal expansion
\begin {equation} 
                                                          \label {1.3}
\ln Z(N,d,g)=N^2\sum_{k=1}^{\infty}\sum_{p=0}^{\infty}
F_{kp}(g)N^{-2p}d^k
\end   {equation} 
Note that $Z(N,d,0)$ is real and therefore $\ln Z(N,d,g)$
is uniquely defined as the formal series over $g$.
We will prove the following theorem.

{\it Theorem. The set of connected vacuum
Feynman diagrams for the model   (\ref
{1.1}), ( \ref {1.2}) in the limit $N\to \infty$ and $d\to 0$ is in one-to-one
correspondence with the set of alternating knot diagrams.
The generating function for the alternating knot diagrams is given by the
expression  }
\begin {equation} 
                                                   \label {1.4}
F(g)=\lim_{d\to 0} \lim_{N\to\infty}\frac{1}{dN^2}
\ln Z(N,d,g)
\end   {equation} 

{\it Remark}. One has the similar proposition for all (not only for alternating)
knot diagrams if one takes the following Lagrangian
\begin {equation} 
                                                          \label {1.5}
S=Tr(A_{\mu}A_{\mu})+Tr(B_{\mu}B_{\mu})+
Tr(A_{\mu}B_{\mu})+\frac{g}{2N}Tr(A_{\mu}
B_{\nu}A_{\mu}B_{\nu})
\end   {equation} 

The large $N$ limit is considered in QCD ,
matrix models and superstring theory \cite {Hof,BIPZ,AAV,AV,BFSS,IKKT,DVV},
the limit $d\to 0$
is considered in the theory of spin glasses
and in polymer physics, see \cite {deG,IA}.

To prove the proposition let us first remind that a
{\it knot} is a smooth embedding
of an oriented circle in oriented
3-space $R^3$. A collection of $k$ pairwise
disjoint knots is called a {\it k-link}.
Two knots are equivalent (have the same isotopy type)
is they are equivalent under a homeomorphism of $R^3$.
\begin{figure}
\begin{center}
\special{em:linewidth 3.4pt}
\unitlength 0.50mm
\linethickness{0.4pt}
\begin{picture}(97.00,148.33)
\emline{50.33}{100.00}{1}{47.96}{94.97}{2}
\emline{47.96}{94.97}{3}{45.74}{90.08}{4}
\emline{45.74}{90.08}{5}{43.66}{85.33}{6}
\emline{43.66}{85.33}{7}{41.74}{80.72}{8}
\emline{41.74}{80.72}{9}{39.97}{76.25}{10}
\emline{39.97}{76.25}{11}{38.35}{71.92}{12}
\emline{38.35}{71.92}{13}{36.88}{67.73}{14}
\emline{36.88}{67.73}{15}{35.56}{63.68}{16}
\emline{35.56}{63.68}{17}{34.39}{59.77}{18}
\emline{34.39}{59.77}{19}{33.37}{55.99}{20}
\emline{33.37}{55.99}{21}{32.50}{52.36}{22}
\emline{32.50}{52.36}{23}{31.79}{48.87}{24}
\emline{31.79}{48.87}{25}{31.22}{45.52}{26}
\emline{31.22}{45.52}{27}{30.80}{42.31}{28}
\emline{30.80}{42.31}{29}{30.54}{39.23}{30}
\emline{30.54}{39.23}{31}{30.42}{36.30}{32}
\emline{30.42}{36.30}{33}{30.45}{33.51}{34}
\emline{30.45}{33.51}{35}{30.64}{30.86}{36}
\emline{30.64}{30.86}{37}{30.97}{28.34}{38}
\emline{30.97}{28.34}{39}{31.46}{25.97}{40}
\emline{31.46}{25.97}{41}{32.10}{23.74}{42}
\emline{32.10}{23.74}{43}{32.88}{21.64}{44}
\emline{32.88}{21.64}{45}{33.82}{19.69}{46}
\emline{33.82}{19.69}{47}{34.91}{17.87}{48}
\emline{34.91}{17.87}{49}{36.15}{16.20}{50}
\emline{36.15}{16.20}{51}{37.53}{14.66}{52}
\emline{37.53}{14.66}{53}{39.07}{13.27}{54}
\emline{39.07}{13.27}{55}{40.76}{12.01}{56}
\emline{40.76}{12.01}{57}{42.60}{10.90}{58}
\emline{42.60}{10.90}{59}{44.59}{9.92}{60}
\emline{44.59}{9.92}{61}{46.73}{9.09}{62}
\emline{46.73}{9.09}{63}{50.67}{8.00}{64}
\emline{50.67}{8.00}{65}{52.97}{8.21}{66}
\emline{52.97}{8.21}{67}{55.14}{8.56}{68}
\emline{55.14}{8.56}{69}{57.20}{9.08}{70}
\emline{57.20}{9.08}{71}{59.13}{9.74}{72}
\emline{59.13}{9.74}{73}{60.93}{10.56}{74}
\emline{60.93}{10.56}{75}{62.62}{11.54}{76}
\emline{62.62}{11.54}{77}{64.18}{12.67}{78}
\emline{64.18}{12.67}{79}{65.63}{13.95}{80}
\emline{65.63}{13.95}{81}{66.94}{15.39}{82}
\emline{66.94}{15.39}{83}{68.14}{16.98}{84}
\emline{68.14}{16.98}{85}{69.22}{18.72}{86}
\emline{69.22}{18.72}{87}{70.17}{20.62}{88}
\emline{70.17}{20.62}{89}{71.00}{22.67}{90}
\emline{71.00}{22.67}{91}{71.71}{24.87}{92}
\emline{71.71}{24.87}{93}{72.29}{27.23}{94}
\emline{72.29}{27.23}{95}{72.76}{29.74}{96}
\emline{72.76}{29.74}{97}{73.10}{32.41}{98}
\emline{73.10}{32.41}{99}{73.32}{35.23}{100}
\emline{73.32}{35.23}{101}{73.41}{38.20}{102}
\emline{73.41}{38.20}{103}{73.39}{41.33}{104}
\emline{73.39}{41.33}{105}{73.24}{44.61}{106}
\emline{73.24}{44.61}{107}{72.97}{48.05}{108}
\emline{72.97}{48.05}{109}{72.58}{51.64}{110}
\emline{72.58}{51.64}{111}{72.07}{55.38}{112}
\emline{72.07}{55.38}{113}{71.43}{59.28}{114}
\emline{71.43}{59.28}{115}{70.67}{63.33}{116}
\emline{43.67}{69.33}{117}{45.56}{67.90}{118}
\emline{45.56}{67.90}{119}{47.41}{66.68}{120}
\emline{47.41}{66.68}{121}{49.23}{65.66}{122}
\emline{49.23}{65.66}{123}{51.02}{64.85}{124}
\emline{51.02}{64.85}{125}{52.77}{64.24}{126}
\emline{52.77}{64.24}{127}{54.48}{63.84}{128}
\emline{54.48}{63.84}{129}{56.16}{63.65}{130}
\emline{56.16}{63.65}{131}{57.81}{63.66}{132}
\emline{57.81}{63.66}{133}{59.42}{63.87}{134}
\emline{59.42}{63.87}{135}{60.99}{64.29}{136}
\emline{60.99}{64.29}{137}{62.53}{64.92}{138}
\emline{62.53}{64.92}{139}{64.03}{65.75}{140}
\emline{64.03}{65.75}{141}{65.50}{66.79}{142}
\emline{65.50}{66.79}{143}{66.94}{68.03}{144}
\emline{66.94}{68.03}{145}{68.33}{69.48}{146}
\emline{68.33}{69.48}{147}{69.70}{71.13}{148}
\emline{69.70}{71.13}{149}{71.03}{72.99}{150}
\emline{71.03}{72.99}{151}{72.32}{75.05}{152}
\emline{72.32}{75.05}{153}{73.58}{77.32}{154}
\emline{73.58}{77.32}{155}{74.80}{79.80}{156}
\emline{74.80}{79.80}{157}{75.99}{82.48}{158}
\emline{75.99}{82.48}{159}{77.14}{85.37}{160}
\emline{77.14}{85.37}{161}{78.33}{88.67}{162}
\emline{78.33}{89.00}{163}{78.78}{92.70}{164}
\emline{78.78}{92.70}{165}{79.15}{96.18}{166}
\emline{79.15}{96.18}{167}{79.41}{99.44}{168}
\emline{79.41}{99.44}{169}{79.59}{102.48}{170}
\emline{79.59}{102.48}{171}{79.67}{105.30}{172}
\emline{79.67}{105.30}{173}{79.66}{107.90}{174}
\emline{79.66}{107.90}{175}{79.56}{110.28}{176}
\emline{79.56}{110.28}{177}{79.36}{112.45}{178}
\emline{79.36}{112.45}{179}{79.07}{114.39}{180}
\emline{79.07}{114.39}{181}{78.68}{116.12}{182}
\emline{78.68}{116.12}{183}{78.21}{117.62}{184}
\emline{78.21}{117.62}{185}{77.63}{118.91}{186}
\emline{77.63}{118.91}{187}{76.97}{119.97}{188}
\emline{76.97}{119.97}{189}{76.21}{120.82}{190}
\emline{76.21}{120.82}{191}{75.36}{121.45}{192}
\emline{75.36}{121.45}{193}{74.42}{121.85}{194}
\emline{74.42}{121.85}{195}{73.38}{122.04}{196}
\emline{73.38}{122.04}{197}{72.25}{122.01}{198}
\emline{72.25}{122.01}{199}{71.03}{121.76}{200}
\emline{71.03}{121.76}{201}{69.71}{121.29}{202}
\emline{69.71}{121.29}{203}{68.30}{120.60}{204}
\emline{68.30}{120.60}{205}{66.80}{119.69}{206}
\emline{66.80}{119.69}{207}{65.20}{118.56}{208}
\emline{65.20}{118.56}{209}{63.52}{117.22}{210}
\emline{63.52}{117.22}{211}{61.73}{115.65}{212}
\emline{61.73}{115.65}{213}{58.67}{112.67}{214}
\emline{67.00}{79.00}{215}{65.02}{83.52}{216}
\emline{65.02}{83.52}{217}{63.10}{87.81}{218}
\emline{63.10}{87.81}{219}{61.23}{91.86}{220}
\emline{61.23}{91.86}{221}{59.42}{95.67}{222}
\emline{59.42}{95.67}{223}{57.67}{99.24}{224}
\emline{57.67}{99.24}{225}{55.98}{102.58}{226}
\emline{55.98}{102.58}{227}{54.35}{105.68}{228}
\emline{54.35}{105.68}{229}{52.77}{108.54}{230}
\emline{52.77}{108.54}{231}{51.25}{111.16}{232}
\emline{51.25}{111.16}{233}{49.79}{113.54}{234}
\emline{49.79}{113.54}{235}{48.38}{115.69}{236}
\emline{48.38}{115.69}{237}{47.04}{117.60}{238}
\emline{47.04}{117.60}{239}{45.75}{119.27}{240}
\emline{45.75}{119.27}{241}{44.52}{120.71}{242}
\emline{44.52}{120.71}{243}{43.34}{121.90}{244}
\emline{43.34}{121.90}{245}{42.22}{122.86}{246}
\emline{42.22}{122.86}{247}{41.16}{123.59}{248}
\emline{41.16}{123.59}{249}{40.16}{124.07}{250}
\emline{40.16}{124.07}{251}{39.22}{124.32}{252}
\emline{39.22}{124.32}{253}{38.33}{124.32}{254}
\emline{38.33}{124.32}{255}{37.50}{124.10}{256}
\emline{37.50}{124.10}{257}{36.73}{123.63}{258}
\emline{36.73}{123.63}{259}{36.02}{122.92}{260}
\emline{36.02}{122.92}{261}{35.36}{121.98}{262}
\emline{35.36}{121.98}{263}{34.76}{120.80}{264}
\emline{34.76}{120.80}{265}{34.22}{119.39}{266}
\emline{34.22}{119.39}{267}{33.74}{117.73}{268}
\emline{33.74}{117.73}{269}{33.31}{115.84}{270}
\emline{33.31}{115.84}{271}{32.67}{111.67}{272}
\emline{32.67}{111.67}{273}{32.49}{110.73}{274}
\emline{32.49}{110.73}{275}{32.38}{109.43}{276}
\emline{32.38}{109.43}{277}{32.34}{107.78}{278}
\emline{32.34}{107.78}{279}{32.36}{105.78}{280}
\emline{32.36}{105.78}{281}{32.44}{103.42}{282}
\emline{32.44}{103.42}{283}{32.59}{100.70}{284}
\emline{32.59}{100.70}{285}{32.81}{97.64}{286}
\emline{32.81}{97.64}{287}{33.09}{94.21}{288}
\emline{33.09}{94.21}{289}{33.67}{88.00}{290}
\put(97.00,68.00){\makebox(0,0)[cc]{$\Longrightarrow$}}
\end{picture}
\special{em:linewidth 0.4pt}
\unitlength 0.50mm
\linethickness{0.4pt}
\begin{picture}(85.00,148.33)
\emline{50.33}{100.00}{1}{47.96}{94.97}{2}
\emline{47.96}{94.97}{3}{45.74}{90.08}{4}
\emline{45.74}{90.08}{5}{43.66}{85.33}{6}
\emline{43.66}{85.33}{7}{41.74}{80.72}{8}
\emline{41.74}{80.72}{9}{39.97}{76.25}{10}
\emline{39.97}{76.25}{11}{38.35}{71.92}{12}
\emline{38.35}{71.92}{13}{36.88}{67.73}{14}
\emline{36.88}{67.73}{15}{35.56}{63.68}{16}
\emline{35.56}{63.68}{17}{34.39}{59.77}{18}
\emline{34.39}{59.77}{19}{33.37}{55.99}{20}
\emline{33.37}{55.99}{21}{32.50}{52.36}{22}
\emline{32.50}{52.36}{23}{31.79}{48.87}{24}
\emline{31.79}{48.87}{25}{31.22}{45.52}{26}
\emline{31.22}{45.52}{27}{30.80}{42.31}{28}
\emline{30.80}{42.31}{29}{30.54}{39.23}{30}
\emline{30.54}{39.23}{31}{30.42}{36.30}{32}
\emline{30.42}{36.30}{33}{30.45}{33.51}{34}
\emline{30.45}{33.51}{35}{30.64}{30.86}{36}
\emline{30.64}{30.86}{37}{30.97}{28.34}{38}
\emline{30.97}{28.34}{39}{31.46}{25.97}{40}
\emline{31.46}{25.97}{41}{32.10}{23.74}{42}
\emline{32.10}{23.74}{43}{32.88}{21.64}{44}
\emline{32.88}{21.64}{45}{33.82}{19.69}{46}
\emline{33.82}{19.69}{47}{34.91}{17.87}{48}
\emline{34.91}{17.87}{49}{36.15}{16.20}{50}
\emline{36.15}{16.20}{51}{37.53}{14.66}{52}
\emline{37.53}{14.66}{53}{39.07}{13.27}{54}
\emline{39.07}{13.27}{55}{40.76}{12.01}{56}
\emline{40.76}{12.01}{57}{42.60}{10.90}{58}
\emline{42.60}{10.90}{59}{44.59}{9.92}{60}
\emline{44.59}{9.92}{61}{46.73}{9.09}{62}
\emline{46.73}{9.09}{63}{50.67}{8.00}{64}
\emline{50.67}{8.00}{65}{52.97}{8.21}{66}
\emline{52.97}{8.21}{67}{55.14}{8.56}{68}
\emline{55.14}{8.56}{69}{57.20}{9.08}{70}
\emline{57.20}{9.08}{71}{59.13}{9.74}{72}
\emline{59.13}{9.74}{73}{60.93}{10.56}{74}
\emline{60.93}{10.56}{75}{62.62}{11.54}{76}
\emline{62.62}{11.54}{77}{64.18}{12.67}{78}
\emline{64.18}{12.67}{79}{65.63}{13.95}{80}
\emline{65.63}{13.95}{81}{66.94}{15.39}{82}
\emline{66.94}{15.39}{83}{68.14}{16.98}{84}
\emline{68.14}{16.98}{85}{69.22}{18.72}{86}
\emline{69.22}{18.72}{87}{70.17}{20.62}{88}
\emline{70.17}{20.62}{89}{71.00}{22.67}{90}
\emline{71.00}{22.67}{91}{71.71}{24.87}{92}
\emline{71.71}{24.87}{93}{72.29}{27.23}{94}
\emline{72.29}{27.23}{95}{72.76}{29.74}{96}
\emline{72.76}{29.74}{97}{73.10}{32.41}{98}
\emline{73.10}{32.41}{99}{73.32}{35.23}{100}
\emline{73.32}{35.23}{101}{73.41}{38.20}{102}
\emline{73.41}{38.20}{103}{73.39}{41.33}{104}
\emline{73.39}{41.33}{105}{73.24}{44.61}{106}
\emline{73.24}{44.61}{107}{72.97}{48.05}{108}
\emline{72.97}{48.05}{109}{72.58}{51.64}{110}
\emline{72.58}{51.64}{111}{72.07}{55.38}{112}
\emline{72.07}{55.38}{113}{71.43}{59.28}{114}
\emline{71.43}{59.28}{115}{70.67}{63.33}{116}
\emline{43.67}{69.33}{117}{45.56}{67.90}{118}
\emline{45.56}{67.90}{119}{47.41}{66.68}{120}
\emline{47.41}{66.68}{121}{49.23}{65.66}{122}
\emline{49.23}{65.66}{123}{51.02}{64.85}{124}
\emline{51.02}{64.85}{125}{52.77}{64.24}{126}
\emline{52.77}{64.24}{127}{54.48}{63.84}{128}
\emline{54.48}{63.84}{129}{56.16}{63.65}{130}
\emline{56.16}{63.65}{131}{57.81}{63.66}{132}
\emline{57.81}{63.66}{133}{59.42}{63.87}{134}
\emline{59.42}{63.87}{135}{60.99}{64.29}{136}
\emline{60.99}{64.29}{137}{62.53}{64.92}{138}
\emline{62.53}{64.92}{139}{64.03}{65.75}{140}
\emline{64.03}{65.75}{141}{65.50}{66.79}{142}
\emline{65.50}{66.79}{143}{66.94}{68.03}{144}
\emline{66.94}{68.03}{145}{68.33}{69.48}{146}
\emline{68.33}{69.48}{147}{69.70}{71.13}{148}
\emline{69.70}{71.13}{149}{71.03}{72.99}{150}
\emline{71.03}{72.99}{151}{72.32}{75.05}{152}
\emline{72.32}{75.05}{153}{73.58}{77.32}{154}
\emline{73.58}{77.32}{155}{74.80}{79.80}{156}
\emline{74.80}{79.80}{157}{75.99}{82.48}{158}
\emline{75.99}{82.48}{159}{77.14}{85.37}{160}
\emline{77.14}{85.37}{161}{78.33}{88.67}{162}
\emline{78.33}{89.00}{163}{78.78}{92.70}{164}
\emline{78.78}{92.70}{165}{79.15}{96.18}{166}
\emline{79.15}{96.18}{167}{79.41}{99.44}{168}
\emline{79.41}{99.44}{169}{79.59}{102.48}{170}
\emline{79.59}{102.48}{171}{79.67}{105.30}{172}
\emline{79.67}{105.30}{173}{79.66}{107.90}{174}
\emline{79.66}{107.90}{175}{79.56}{110.28}{176}
\emline{79.56}{110.28}{177}{79.36}{112.45}{178}
\emline{79.36}{112.45}{179}{79.07}{114.39}{180}
\emline{79.07}{114.39}{181}{78.68}{116.12}{182}
\emline{78.68}{116.12}{183}{78.21}{117.62}{184}
\emline{78.21}{117.62}{185}{77.63}{118.91}{186}
\emline{77.63}{118.91}{187}{76.97}{119.97}{188}
\emline{76.97}{119.97}{189}{76.21}{120.82}{190}
\emline{76.21}{120.82}{191}{75.36}{121.45}{192}
\emline{75.36}{121.45}{193}{74.42}{121.85}{194}
\emline{74.42}{121.85}{195}{73.38}{122.04}{196}
\emline{73.38}{122.04}{197}{72.25}{122.01}{198}
\emline{72.25}{122.01}{199}{71.03}{121.76}{200}
\emline{71.03}{121.76}{201}{69.71}{121.29}{202}
\emline{69.71}{121.29}{203}{68.30}{120.60}{204}
\emline{68.30}{120.60}{205}{66.80}{119.69}{206}
\emline{66.80}{119.69}{207}{65.20}{118.56}{208}
\emline{65.20}{118.56}{209}{63.52}{117.22}{210}
\emline{63.52}{117.22}{211}{61.73}{115.65}{212}
\emline{61.73}{115.65}{213}{58.67}{112.67}{214}
\emline{67.00}{79.00}{215}{65.02}{83.52}{216}
\emline{65.02}{83.52}{217}{63.10}{87.81}{218}
\emline{63.10}{87.81}{219}{61.23}{91.86}{220}
\emline{61.23}{91.86}{221}{59.42}{95.67}{222}
\emline{59.42}{95.67}{223}{57.67}{99.24}{224}
\emline{57.67}{99.24}{225}{55.98}{102.58}{226}
\emline{55.98}{102.58}{227}{54.35}{105.68}{228}
\emline{54.35}{105.68}{229}{52.77}{108.54}{230}
\emline{52.77}{108.54}{231}{51.25}{111.16}{232}
\emline{51.25}{111.16}{233}{49.79}{113.54}{234}
\emline{49.79}{113.54}{235}{48.38}{115.69}{236}
\emline{48.38}{115.69}{237}{47.04}{117.60}{238}
\emline{47.04}{117.60}{239}{45.75}{119.27}{240}
\emline{45.75}{119.27}{241}{44.52}{120.71}{242}
\emline{44.52}{120.71}{243}{43.34}{121.90}{244}
\emline{43.34}{121.90}{245}{42.22}{122.86}{246}
\emline{42.22}{122.86}{247}{41.16}{123.59}{248}
\emline{41.16}{123.59}{249}{40.16}{124.07}{250}
\emline{40.16}{124.07}{251}{39.22}{124.32}{252}
\emline{39.22}{124.32}{253}{38.33}{124.32}{254}
\emline{38.33}{124.32}{255}{37.50}{124.10}{256}
\emline{37.50}{124.10}{257}{36.73}{123.63}{258}
\emline{36.73}{123.63}{259}{36.02}{122.92}{260}
\emline{36.02}{122.92}{261}{35.36}{121.98}{262}
\emline{35.36}{121.98}{263}{34.76}{120.80}{264}
\emline{34.76}{120.80}{265}{34.22}{119.39}{266}
\emline{34.22}{119.39}{267}{33.74}{117.73}{268}
\emline{33.74}{117.73}{269}{33.31}{115.84}{270}
\emline{33.31}{115.84}{271}{32.67}{111.67}{272}
\emline{32.67}{111.67}{273}{32.49}{110.73}{274}
\emline{32.49}{110.73}{275}{32.38}{109.43}{276}
\emline{32.38}{109.43}{277}{32.34}{107.78}{278}
\emline{32.34}{107.78}{279}{32.36}{105.78}{280}
\emline{32.36}{105.78}{281}{32.44}{103.42}{282}
\emline{32.44}{103.42}{283}{32.59}{100.70}{284}
\emline{32.59}{100.70}{285}{32.81}{97.64}{286}
\emline{32.81}{97.64}{287}{33.09}{94.21}{288}
\emline{33.09}{94.21}{289}{33.67}{88.00}{290}
\emline{43.33}{69.67}{291}{41.47}{71.31}{292}
\emline{41.47}{71.31}{293}{39.81}{73.06}{294}
\emline{39.81}{73.06}{295}{38.35}{74.90}{296}
\emline{38.35}{74.90}{297}{37.11}{76.86}{298}
\emline{37.11}{76.86}{299}{36.07}{78.91}{300}
\emline{36.07}{78.91}{301}{35.23}{81.07}{302}
\emline{35.23}{81.07}{303}{34.61}{83.33}{304}
\emline{34.61}{83.33}{305}{34.00}{87.67}{306}
\emline{50.67}{100.00}{307}{52.64}{103.58}{308}
\emline{52.64}{103.58}{309}{54.33}{106.56}{310}
\emline{54.33}{106.56}{311}{55.75}{108.92}{312}
\emline{55.75}{108.92}{313}{56.89}{110.67}{314}
\emline{56.89}{110.67}{315}{57.75}{111.81}{316}
\emline{57.75}{111.81}{317}{58.33}{112.33}{318}
\emline{71.00}{63.67}{319}{70.03}{67.82}{320}
\emline{70.03}{67.82}{321}{69.21}{71.32}{322}
\emline{69.21}{71.32}{323}{68.53}{74.17}{324}
\emline{68.53}{74.17}{325}{68.01}{76.38}{326}
\emline{68.01}{76.38}{327}{67.62}{77.92}{328}
\emline{67.62}{77.92}{329}{67.33}{79.00}{330}
\put(31.00,79.00){\makebox(0,0)[cc]{$-$}}
\put(46.67,63.67){\makebox(0,0)[cc]{$-$}}
\put(28.33,56.67){\makebox(0,0)[cc]{$+$}}
\put(41.00,92.67){\makebox(0,0)[cc]{$+$}}
\put(46.33,104.33){\makebox(0,0)[cc]{$-$}}
\put(55.33,118.00){\makebox(0,0)[cc]{$-$}}
\put(49.33,125.33){\makebox(0,0)[cc]{$+$}}
\put(61.67,103.67){\makebox(0,0)[cc]{$+$}}
\put(79.00,72.33){\makebox(0,0)[cc]{$+$}}
\put(60.67,55.67){\makebox(0,0)[cc]{$+$}}
\put(62.67,74.00){\makebox(0,0)[cc]{$-$}}
\put(81.33,53.00){\makebox(0,0)[cc]{$-$}}
\end{picture}
\end{center}
\caption{Trefoil}\label{fig1}
\end{figure}

A knot $K$ can be represented
by a regular projection ${\tilde K}$  onto the plane having at most a
finite number of transverse double points. For the plane curve ${\tilde K}$
one has to indicate which line is up $(+)$ and which line is down $(-)$
in an intersection point, see Fig. 1. In this way we get a graph
on the plane which has 4 legs in each vertex and also has the $(+ ~-)$
prescription. This graph is called the {\it knot diagram}.
A knot diagram is called alternating if it has alternating
$+$ and $-$ along a line. Two knot
diagrams are called the Reidemeister equivalent if they define
equivalent knotes. Reidemeister equivalence is generated by the three
moves that are illustrated in Fig 2.
\begin{figure}
\begin{center}
\special{em:linewidth 2.4pt}
\unitlength 0.50mm
\linethickness{0.4pt}
\begin{picture}(180.00,107.08)
\emline{9.67}{43.67}{1}{11.36}{41.59}{2}
\emline{11.36}{41.59}{3}{13.08}{39.69}{4}
\emline{13.08}{39.69}{5}{14.84}{37.97}{6}
\emline{14.84}{37.97}{7}{16.63}{36.43}{8}
\emline{16.63}{36.43}{9}{18.45}{35.07}{10}
\emline{18.45}{35.07}{11}{20.31}{33.90}{12}
\emline{20.31}{33.90}{13}{22.20}{32.91}{14}
\emline{22.20}{32.91}{15}{24.13}{32.09}{16}
\emline{24.13}{32.09}{17}{26.09}{31.47}{18}
\emline{26.09}{31.47}{19}{28.08}{31.02}{20}
\emline{28.08}{31.02}{21}{30.11}{30.75}{22}
\emline{30.11}{30.75}{23}{32.17}{30.67}{24}
\emline{32.17}{30.67}{25}{34.27}{30.77}{26}
\emline{34.27}{30.77}{27}{36.39}{31.05}{28}
\emline{36.39}{31.05}{29}{38.56}{31.51}{30}
\emline{38.56}{31.51}{31}{40.75}{32.15}{32}
\emline{40.75}{32.15}{33}{42.98}{32.98}{34}
\emline{42.98}{32.98}{35}{45.25}{33.99}{36}
\emline{45.25}{33.99}{37}{47.54}{35.18}{38}
\emline{47.54}{35.18}{39}{49.87}{36.55}{40}
\emline{49.87}{36.55}{41}{52.24}{38.10}{42}
\emline{52.24}{38.10}{43}{54.64}{39.84}{44}
\emline{54.64}{39.84}{45}{57.07}{41.75}{46}
\emline{57.07}{41.75}{47}{59.33}{43.67}{48}
\emline{8.33}{24.00}{49}{10.71}{25.60}{50}
\emline{10.71}{25.60}{51}{12.98}{27.06}{52}
\emline{12.98}{27.06}{53}{15.14}{28.38}{54}
\emline{15.14}{28.38}{55}{18.00}{30.00}{56}
\emline{26.33}{34.67}{57}{28.33}{35.74}{58}
\emline{28.33}{35.74}{59}{30.37}{36.81}{60}
\emline{30.37}{36.81}{61}{32.44}{37.86}{62}
\emline{32.44}{37.86}{63}{34.55}{38.90}{64}
\emline{34.55}{38.90}{65}{36.70}{39.92}{66}
\emline{36.70}{39.92}{67}{38.88}{40.94}{68}
\emline{38.88}{40.94}{69}{41.10}{41.94}{70}
\emline{41.10}{41.94}{71}{43.36}{42.93}{72}
\emline{43.36}{42.93}{73}{45.65}{43.90}{74}
\emline{45.65}{43.90}{75}{47.98}{44.87}{76}
\emline{47.98}{44.87}{77}{50.35}{45.82}{78}
\emline{50.35}{45.82}{79}{52.75}{46.76}{80}
\emline{52.75}{46.76}{81}{55.19}{47.69}{82}
\emline{55.19}{47.69}{83}{59.67}{49.33}{84}
\emline{10.00}{55.67}{85}{12.87}{54.88}{86}
\emline{12.87}{54.88}{87}{15.64}{54.07}{88}
\emline{15.64}{54.07}{89}{18.31}{53.25}{90}
\emline{18.31}{53.25}{91}{20.89}{52.41}{92}
\emline{20.89}{52.41}{93}{23.36}{51.56}{94}
\emline{23.36}{51.56}{95}{25.74}{50.69}{96}
\emline{25.74}{50.69}{97}{28.02}{49.80}{98}
\emline{28.02}{49.80}{99}{30.21}{48.90}{100}
\emline{30.21}{48.90}{101}{32.29}{47.98}{102}
\emline{32.29}{47.98}{103}{34.28}{47.04}{104}
\emline{34.28}{47.04}{105}{36.17}{46.09}{106}
\emline{36.17}{46.09}{107}{39.33}{44.33}{108}
\emline{45.67}{41.00}{109}{47.34}{39.92}{110}
\emline{47.34}{39.92}{111}{49.67}{38.67}{112}
\emline{53.33}{35.67}{113}{55.00}{34.33}{114}
\emline{55.00}{34.33}{115}{56.67}{32.67}{116}
\emline{56.67}{32.67}{117}{58.33}{30.66}{118}
\emline{58.33}{30.66}{119}{60.00}{28.33}{120}
\emline{85.33}{23.00}{121}{87.34}{26.12}{122}
\emline{87.34}{26.12}{123}{89.34}{29.01}{124}
\emline{89.34}{29.01}{125}{91.32}{31.69}{126}
\emline{91.32}{31.69}{127}{93.28}{34.14}{128}
\emline{93.28}{34.14}{129}{95.22}{36.38}{130}
\emline{95.22}{36.38}{131}{97.15}{38.39}{132}
\emline{97.15}{38.39}{133}{99.06}{40.18}{134}
\emline{99.06}{40.18}{135}{100.95}{41.75}{136}
\emline{100.95}{41.75}{137}{102.82}{43.10}{138}
\emline{102.82}{43.10}{139}{104.67}{44.23}{140}
\emline{104.67}{44.23}{141}{106.51}{45.14}{142}
\emline{106.51}{45.14}{143}{108.33}{45.83}{144}
\emline{108.33}{45.83}{145}{110.13}{46.30}{146}
\emline{110.13}{46.30}{147}{111.91}{46.54}{148}
\emline{111.91}{46.54}{149}{113.68}{46.57}{150}
\emline{113.68}{46.57}{151}{115.42}{46.37}{152}
\emline{115.42}{46.37}{153}{117.15}{45.95}{154}
\emline{117.15}{45.95}{155}{118.86}{45.32}{156}
\emline{118.86}{45.32}{157}{120.56}{44.46}{158}
\emline{120.56}{44.46}{159}{122.23}{43.38}{160}
\emline{122.23}{43.38}{161}{123.89}{42.08}{162}
\emline{123.89}{42.08}{163}{125.53}{40.55}{164}
\emline{125.53}{40.55}{165}{127.15}{38.81}{166}
\emline{127.15}{38.81}{167}{128.76}{36.85}{168}
\emline{128.76}{36.85}{169}{130.34}{34.66}{170}
\emline{130.34}{34.66}{171}{131.91}{32.26}{172}
\emline{131.91}{32.26}{173}{133.46}{29.63}{174}
\emline{133.46}{29.63}{175}{134.99}{26.79}{176}
\emline{134.99}{26.79}{177}{137.00}{22.67}{178}
\emline{85.00}{13.00}{179}{88.43}{13.39}{180}
\emline{88.43}{13.39}{181}{91.68}{13.84}{182}
\emline{91.68}{13.84}{183}{94.74}{14.33}{184}
\emline{94.74}{14.33}{185}{97.63}{14.87}{186}
\emline{97.63}{14.87}{187}{100.34}{15.46}{188}
\emline{100.34}{15.46}{189}{102.87}{16.09}{190}
\emline{102.87}{16.09}{191}{105.22}{16.77}{192}
\emline{105.22}{16.77}{193}{107.39}{17.51}{194}
\emline{107.39}{17.51}{195}{109.37}{18.29}{196}
\emline{109.37}{18.29}{197}{111.18}{19.11}{198}
\emline{111.18}{19.11}{199}{112.81}{19.99}{200}
\emline{112.81}{19.99}{201}{114.26}{20.92}{202}
\emline{114.26}{20.92}{203}{115.53}{21.89}{204}
\emline{115.53}{21.89}{205}{117.00}{23.33}{206}
\emline{117.00}{23.33}{207}{119.31}{26.11}{208}
\emline{119.31}{26.11}{209}{121.25}{28.61}{210}
\emline{121.25}{28.61}{211}{122.81}{30.84}{212}
\emline{122.81}{30.84}{213}{124.00}{32.79}{214}
\emline{124.00}{32.79}{215}{124.82}{34.46}{216}
\emline{124.82}{34.46}{217}{125.33}{36.33}{218}
\emline{127.67}{43.00}{219}{128.27}{44.88}{220}
\emline{128.27}{44.88}{221}{129.19}{46.48}{222}
\emline{129.19}{46.48}{223}{130.44}{47.81}{224}
\emline{130.44}{47.81}{225}{132.00}{48.85}{226}
\emline{132.00}{48.85}{227}{133.90}{49.63}{228}
\emline{133.90}{49.63}{229}{136.11}{50.12}{230}
\emline{136.11}{50.12}{231}{140.33}{50.33}{232}
\emline{85.00}{51.67}{233}{86.12}{48.96}{234}
\emline{86.12}{48.96}{235}{87.30}{46.53}{236}
\emline{87.30}{46.53}{237}{88.56}{44.37}{238}
\emline{88.56}{44.37}{239}{89.89}{42.49}{240}
\emline{89.89}{42.49}{241}{91.29}{40.88}{242}
\emline{91.29}{40.88}{243}{94.00}{38.67}{244}
\emline{121.33}{20.33}{245}{123.33}{19.15}{246}
\emline{123.33}{19.15}{247}{125.42}{18.00}{248}
\emline{125.42}{18.00}{249}{127.63}{16.90}{250}
\emline{127.63}{16.90}{251}{129.93}{15.83}{252}
\emline{129.93}{15.83}{253}{132.34}{14.80}{254}
\emline{132.34}{14.80}{255}{134.85}{13.81}{256}
\emline{134.85}{13.81}{257}{138.00}{12.67}{258}
\emline{99.33}{34.00}{259}{101.52}{32.39}{260}
\emline{101.52}{32.39}{261}{103.71}{30.90}{262}
\emline{103.71}{30.90}{263}{105.89}{29.52}{264}
\emline{105.89}{29.52}{265}{108.08}{28.25}{266}
\emline{108.08}{28.25}{267}{110.27}{27.09}{268}
\emline{110.27}{27.09}{269}{113.33}{25.67}{270}
\put(73.33,34.67){\makebox(0,0)[cc]{$\Longleftrightarrow$}}
\emline{15.67}{71.00}{271}{15.67}{101.00}{272}
\emline{38.00}{71.00}{273}{38.00}{77.67}{274}
\emline{38.00}{77.80}{275}{39.00}{81.47}{276}
\emline{38.33}{101.00}{277}{38.33}{92.00}{278}
\emline{38.33}{92.00}{279}{39.19}{89.81}{280}
\emline{39.19}{89.81}{281}{40.15}{87.96}{282}
\emline{40.15}{87.96}{283}{41.21}{86.45}{284}
\emline{41.21}{86.45}{285}{42.38}{85.27}{286}
\emline{42.38}{85.27}{287}{43.66}{84.43}{288}
\emline{43.66}{84.43}{289}{45.03}{83.92}{290}
\emline{45.03}{83.92}{291}{48.10}{83.93}{292}
\emline{48.10}{83.93}{293}{49.79}{84.43}{294}
\emline{49.79}{84.43}{295}{51.58}{85.28}{296}
\emline{51.58}{85.28}{297}{54.67}{87.33}{298}
\emline{54.67}{87.33}{299}{55.92}{89.58}{300}
\emline{55.92}{89.58}{301}{56.61}{91.70}{302}
\emline{56.61}{91.70}{303}{56.73}{93.70}{304}
\emline{56.73}{93.70}{305}{56.29}{95.58}{306}
\emline{56.29}{95.58}{307}{55.00}{97.67}{308}
\emline{55.00}{97.67}{309}{53.28}{98.77}{310}
\emline{53.28}{98.77}{311}{51.60}{99.40}{312}
\emline{51.60}{99.40}{313}{49.98}{99.56}{314}
\emline{49.98}{99.56}{315}{48.42}{99.25}{316}
\emline{48.42}{99.25}{317}{46.90}{98.47}{318}
\emline{46.90}{98.47}{319}{45.44}{97.23}{320}
\emline{45.44}{97.23}{321}{44.03}{95.51}{322}
\emline{44.03}{95.51}{323}{42.67}{93.33}{324}
\put(26.33,86.67){\makebox(0,0)[cc]{$\Longleftrightarrow$}}
\emline{79.67}{100.33}{325}{81.08}{98.17}{326}
\emline{81.08}{98.17}{327}{82.52}{96.20}{328}
\emline{82.52}{96.20}{329}{83.97}{94.44}{330}
\emline{83.97}{94.44}{331}{85.45}{92.88}{332}
\emline{85.45}{92.88}{333}{86.96}{91.52}{334}
\emline{86.96}{91.52}{335}{88.48}{90.36}{336}
\emline{88.48}{90.36}{337}{90.03}{89.40}{338}
\emline{90.03}{89.40}{339}{91.60}{88.65}{340}
\emline{91.60}{88.65}{341}{93.20}{88.09}{342}
\emline{93.20}{88.09}{343}{94.81}{87.74}{344}
\emline{94.81}{87.74}{345}{96.45}{87.59}{346}
\emline{96.45}{87.59}{347}{98.11}{87.64}{348}
\emline{98.11}{87.64}{349}{99.80}{87.89}{350}
\emline{99.80}{87.89}{351}{101.51}{88.34}{352}
\emline{101.51}{88.34}{353}{103.24}{88.99}{354}
\emline{103.24}{88.99}{355}{104.99}{89.85}{356}
\emline{104.99}{89.85}{357}{106.76}{90.91}{358}
\emline{106.76}{90.91}{359}{108.56}{92.16}{360}
\emline{108.56}{92.16}{361}{110.38}{93.62}{362}
\emline{110.38}{93.62}{363}{112.23}{95.28}{364}
\emline{112.23}{95.28}{365}{114.09}{97.14}{366}
\emline{114.09}{97.14}{367}{116.67}{100.00}{368}
\emline{79.67}{79.00}{369}{81.50}{82.48}{370}
\emline{81.50}{82.48}{371}{83.09}{85.16}{372}
\emline{83.09}{85.16}{373}{84.46}{87.02}{374}
\emline{84.46}{87.02}{375}{86.33}{88.33}{376}
\emline{91.33}{91.67}{377}{93.16}{92.96}{378}
\emline{93.16}{92.96}{379}{95.07}{93.74}{380}
\emline{95.07}{93.74}{381}{99.19}{93.74}{382}
\emline{99.19}{93.74}{383}{101.38}{92.96}{384}
\emline{101.38}{92.96}{385}{103.67}{91.67}{386}
\emline{108.33}{88.33}{387}{110.23}{86.82}{388}
\emline{110.23}{86.82}{389}{111.87}{85.13}{390}
\emline{111.87}{85.13}{391}{113.26}{83.28}{392}
\emline{113.26}{83.28}{393}{114.39}{81.26}{394}
\emline{114.39}{81.26}{395}{115.67}{77.67}{396}
\emline{140.00}{76.67}{397}{141.69}{79.09}{398}
\emline{141.69}{79.09}{399}{143.38}{81.30}{400}
\emline{143.38}{81.30}{401}{145.06}{83.29}{402}
\emline{145.06}{83.29}{403}{146.74}{85.07}{404}
\emline{146.74}{85.07}{405}{148.41}{86.64}{406}
\emline{148.41}{86.64}{407}{150.07}{88.00}{408}
\emline{150.07}{88.00}{409}{151.73}{89.14}{410}
\emline{151.73}{89.14}{411}{153.38}{90.07}{412}
\emline{153.38}{90.07}{413}{155.03}{90.79}{414}
\emline{155.03}{90.79}{415}{156.67}{91.29}{416}
\emline{156.67}{91.29}{417}{158.30}{91.59}{418}
\emline{158.30}{91.59}{419}{159.93}{91.67}{420}
\emline{159.93}{91.67}{421}{161.56}{91.53}{422}
\emline{161.56}{91.53}{423}{163.18}{91.19}{424}
\emline{163.18}{91.19}{425}{164.79}{90.63}{426}
\emline{164.79}{90.63}{427}{166.39}{89.86}{428}
\emline{166.39}{89.86}{429}{168.00}{88.88}{430}
\emline{168.00}{88.88}{431}{169.59}{87.68}{432}
\emline{169.59}{87.68}{433}{171.18}{86.28}{434}
\emline{171.18}{86.28}{435}{172.76}{84.65}{436}
\emline{172.76}{84.65}{437}{174.34}{82.82}{438}
\emline{174.34}{82.82}{439}{175.91}{80.78}{440}
\emline{175.91}{80.78}{441}{177.48}{78.52}{442}
\emline{177.48}{78.52}{443}{179.17}{75.83}{444}
\emline{140.00}{100.00}{445}{140.37}{97.30}{446}
\emline{140.37}{97.30}{447}{140.92}{94.77}{448}
\emline{140.92}{94.77}{449}{141.67}{92.40}{450}
\emline{141.67}{92.40}{451}{142.59}{90.19}{452}
\emline{142.59}{90.19}{453}{143.70}{88.14}{454}
\emline{143.70}{88.14}{455}{145.00}{86.25}{456}
\emline{149.58}{83.33}{457}{151.64}{81.71}{458}
\emline{151.64}{81.71}{459}{153.65}{80.41}{460}
\emline{153.65}{80.41}{461}{155.61}{79.42}{462}
\emline{155.61}{79.42}{463}{157.51}{78.75}{464}
\emline{157.51}{78.75}{465}{159.37}{78.39}{466}
\emline{159.37}{78.39}{467}{161.18}{78.34}{468}
\emline{161.18}{78.34}{469}{162.94}{78.61}{470}
\emline{162.94}{78.61}{471}{164.65}{79.20}{472}
\emline{164.65}{79.20}{473}{166.31}{80.09}{474}
\emline{166.31}{80.09}{475}{169.17}{82.50}{476}
\emline{175.83}{87.50}{477}{176.95}{89.08}{478}
\emline{176.95}{89.08}{479}{177.85}{90.87}{480}
\emline{177.85}{90.87}{481}{178.54}{92.88}{482}
\emline{178.54}{92.88}{483}{179.01}{95.11}{484}
\emline{179.01}{95.11}{485}{179.26}{97.56}{486}
\emline{179.26}{97.56}{487}{179.17}{102.50}{488}
\put(130.92,86.25){\makebox(0,0)[cc]{$\Longleftrightarrow$}}
\end{picture}
\end{center}
\caption{Reidemeister moves}\label{fig2}
\end{figure}
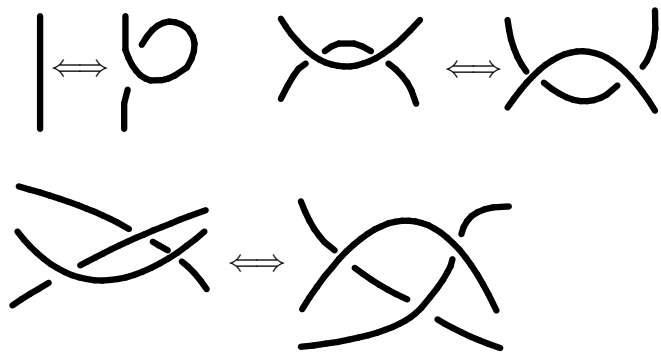

The fundamental group  $\pi (R^3\backslash K)$
is called the group of the knot $K$.
The knot group is generated by simbols corresponding to legs
of the knot diagram subject to some relations.

Now let us consider the integral (\ref {1.1}). We obtain the Feynman
diagram technique by expanding (\ref {1.1}) into the formal perturbation
series over the coupling constant $g$ and computing the corresponding
Gaussian integrals
\begin {equation} 
                                                          \label {1.6}
Z=\sum_{k=0}^{\infty}\frac{1}{k!}(\frac{ig}{N})^k
\int (Tr(A_{\mu}
B_{\nu}A_{\mu}B_{\nu}))^k e^{iTr(A_{\mu}B_{\mu})}dAdB
\end   {equation} 
We have the following propagators  and the vertex function (see Fig.3)
\begin{figure}
\begin{center}
\unitlength .80mm
\linethickness{0.4pt}
\begin{picture}(160.83,71.33)
\put(15.33,60.00){\line(1,0){30.00}}
\put(45.33,64.67){\line(-1,0){29.66}}
\put(44.33,60.00){\vector(1,0){0.2}}
\put(15.67,60.00){\line(1,0){28.66}}
\put(15.67,64.67){\vector(-1,0){0.2}}
\put(44.67,64.67){\line(-1,0){29.00}}
\put(16.00,62.33){\line(1,0){29.00}}
\put(13.00,70.67){\makebox(0,0)[cc]{$i$}}
\put(6.00,62.33){\makebox(0,0)[cc]{$\mu$}}
\put(10.33,61.00){\makebox(0,0)[cc]{{\small $+$}}}
\put(12.67,53.67){\makebox(0,0)[cc]{$j$}}
\put(49.67,71.33){\makebox(0,0)[cc]{$n$}}
\put(48.67,61.33){\makebox(0,0)[cc]{{\small $-$}}}
\put(55.34,62.00){\makebox(0,0)[cc]{$\nu$}}
\put(49.67,54.00){\makebox(0,0)[cc]{$m$}}
\put(62.00,62.67){\makebox(0,0)[cc]{$=$}}
\put(64.66,62.34){\makebox(0,0)[lc]
{$i\delta _{\mu\nu}\delta _{in}\delta _{jm}$}}
\put(110.42,59.67){\line(1,0){30.00}}
\put(140.42,64.34){\line(-1,0){29.66}}
\put(139.42,59.67){\vector(1,0){0.2}}
\put(110.76,59.67){\line(1,0){28.66}}
\put(110.76,64.34){\vector(-1,0){0.2}}
\put(139.76,64.34){\line(-1,0){29.00}}
\put(111.09,62.00){\line(1,0){29.00}}
\put(108.09,70.34){\makebox(0,0)[cc]{$i$}}
\put(101.09,62.00){\makebox(0,0)[cc]{$\mu$}}
\put(105.42,60.67){\makebox(0,0)[cc]{{\small $+$}}}
\put(107.76,53.34){\makebox(0,0)[cc]{$j$}}
\put(144.76,71.00){\makebox(0,0)[cc]{$n$}}
\put(143.75,61.00){\makebox(0,0)[cc]{{\small $+$}}}
\put(150.42,61.67){\makebox(0,0)[cc]{$\nu$}}
\put(144.76,53.67){\makebox(0,0)[cc]{$m$}}
\put(157.17,61.58){\makebox(0,0)[cc]{$=$}}
\put(160.83,62.25){\makebox(0,0)[lc]{$0$}}
\put(15.00,25.00){\line(1,0){40.33}}
\put(15.33,22.33){\line(1,0){16.67}}
\put(32.00,22.33){\line(0,-1){15.66}}
\put(34.67,7.00){\line(0,1){33.33}}
\put(37.67,6.67){\line(0,1){15.66}}
\put(37.67,22.33){\line(1,0){18.00}}
\put(55.33,28.33){\line(-1,0){17.33}}
\put(38.00,28.33){\line(0,1){13.34}}
\put(32.00,42.00){\line(0,-1){13.67}}
\put(32.00,28.33){\line(-1,0){16.67}}
\put(12.67,31.33){\makebox(0,0)[cc]{$i$}}
\put(12.67,18.33){\makebox(0,0)[cc]{$q$}}
\put(10.00,23.67){\makebox(0,0)[cc]{$+$}}
\put(4.67,26.67){\makebox(0,0)[cc]{$\mu$}}
\put(27.67,43.67){\makebox(0,0)[cc]{$j$}}
\put(36.00,43.67){\makebox(0,0)[cc]{$-$}}
\put(36.00,48.33){\makebox(0,0)[cc]{$\nu$}}
\put(42.00,44.00){\makebox(0,0)[cc]{$k$}}
\put(56.67,32.00){\makebox(0,0)[cc]{$l$}}
\put(61.67,28.33){\makebox(0,0)[cc]{$\alpha$}}
\put(59.67,23.00){\makebox(0,0)[cc]{$+$}}
\put(27.33,3.33){\makebox(0,0)[cc]{$p$}}
\put(33.00,1.00){\makebox(0,0)[cc]{$\beta$}}
\put(36.33,3.67){\makebox(0,0)[cc]{$-$}}
\put(41.67,3.33){\makebox(0,0)[cc]{$n$}}
\put(70.00,25.67){\makebox(0,0)[cc]{$=$}}
\put(79.00,25.67){\makebox(0,0)[lc]
{$\frac{ig}{N}\delta _{\mu\alpha}\delta _{\nu\beta}
\delta _{ij}\delta_{kl}\delta_{mn}\delta_{pq}$}}
\end{picture}
\end{center}
\caption{Elements of Feynman diagrams }\label{fig3}
\end{figure}
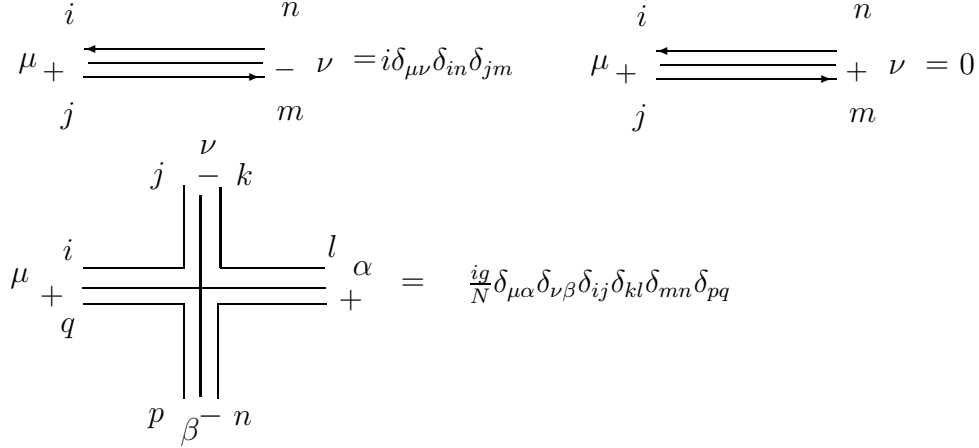

\begin {equation} 
                                                          \label {1.7}
<A_{\mu}^{ kl}B_{\nu }^{mn}>=i \delta_{\mu\nu}\delta^{kn}
\delta^{lm}
\end   {equation} 
$$
<A_{\mu}^{ kl}B_{\nu }^{mn}>=0, ~~~<A_{\mu}^{ kl}B_{\nu}^{ mn}>= 0
$$
where
$$
<X>=\int X e^{iTr(A_{\mu}B_{\mu})}dAdB /
\int e^{iTr(A_{\mu}B_{\mu})}dAdB
$$
The propagators are represented by triple lines each one corresponding
to the separate propagation of its two indices.  The middle line carries
a Greek index $\mu,\nu,...$ and all others carry Latin indices.
The matrix $A_{\mu}$ corresponds to $+$ in Fig.3 and the matrix $B_{\mu}$
corresponds to $-$.
To compute a contribution to the partition function $Z$ from the $n-$th
order of perturbation theory we have to draw all diagrams with $n$ vertices,
see Fig. 4 for $n=3$.
\begin{figure}
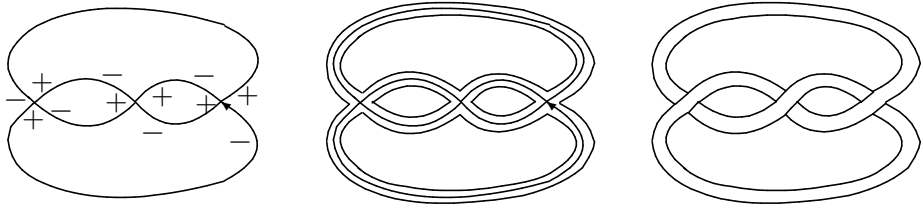

\begin{center}
\special{em:linewidth 0.6pt}
\unitlength 0.5mm
\linethickness{0.6pt}

\end{center}
\caption{Third order graph (trefoil) in triple line
representation}\label{fig3a} \end{figure}

The known connection between planarity and the large $N$ limit \cite {Hof}
is based on the Euler theorem. A general Feynman
diagram consists of $L$ lines
(propagators), $V$ vertices and $C$ closed loops of Latin indices.
The contribution of the diagram is proportional to $(g/N)^VN^C=
g^VN^{C-V}$. For a connected diagram one has $L=2V$. Each closed loop
of Latin index may be considered as a face of a polyhedron
and the Euler relation reads $V-L+C=2-2p$ where $p=0,1,...$
is the number of holes of the surface on which the polyhedron
is drawn (genus of the Riemannian surface). Therefore $C-V=2-2p$
and the contribution of the diagram is proportional to $g^VN^{2-2p}$.
This justifies the factor $N^{2-2p}$ in (\ref {1.3}).  We obtain that
the principal contribution comes from the planar diagrams with $p=0$.
Now in the similar vine one gets that if a planar diagram
has $k$ closed loops with Greek indices (i.e. one has a $k$-link)
then the contribution
of the diagram is proportional to $d^k$. In particular the contribution
of the knot diagrams $( k=1 ) $ is proportional to $d$. This proves the
relations (\ref {1.3}), ( \ref {1.4}).

{\it Remark}. It is interesting that the first
Reidemeister move in Fig. 3 admits
a natural interpretation in the Feynman diagram technique
for the Lagrangian (\ref {1.2}). This move generates
the so called tadpole diagrams and can be removed by
using the Wick normal product if we will use the Lagrangian
in the normal form
\begin {equation} 
                                                          \label {1.8}
S=Tr(A_{\mu}B_{\mu})+\frac{g}{2N}:Tr(A_{\mu}
B_{\nu}
A_{\mu}B_{\nu}):
\end   {equation} 
where
\begin {equation} 
                                                          \label {1.9}
:Tr(A_{\mu}B_{\nu}A_{\mu}B_{\nu}): =
Tr(A_{\mu}B_{\nu}A_{\mu}B_{\nu})-2dNTr(A_{\mu}B_{\mu}) -2dN^2
\end   {equation} 
The lowest order diagrams in matrix theory and corresponding
knot diagrams are presented in Fig.6. Up to the 8-th order
there is one-to-one correspondence between the matrix diagrams
describing the limit $N\to\infty, ~d\to 0$ and non-isotopic
knot diagrams. At the 8-th order there are 3 non-alternating
knot diagrams and to reproduce them we have to consider the
matrix theory  with the Lagrangian (\ref {1.5}).

\newpage
\begin{figure}[here]
\begin{center}
\special{em:linewidth 0.4pt}
\unitlength 0.5mm
\linethickness{0.4pt}
\begin{picture}(28.40,40.00)
\emline{13.00}{9.67}{1}{13.00}{40.00}{2}
\emline{20.67}{33.06}{3}{22.40}{32.87}{4}
\emline{22.40}{32.87}{5}{24.04}{32.29}{6}
\emline{24.04}{32.29}{7}{25.51}{31.36}{8}
\emline{25.51}{31.36}{9}{26.73}{30.13}{10}
\emline{26.73}{30.13}{11}{27.65}{28.65}{12}
\emline{27.65}{28.65}{13}{28.22}{27.00}{14}
\emline{28.22}{27.00}{15}{28.40}{25.27}{16}
\emline{28.40}{25.27}{17}{28.19}{23.55}{18}
\emline{28.19}{23.55}{19}{27.60}{21.91}{20}
\emline{27.60}{21.91}{21}{26.66}{20.45}{22}
\emline{26.66}{20.45}{23}{25.42}{19.23}{24}
\emline{25.42}{19.23}{25}{23.93}{18.32}{26}
\emline{23.93}{18.32}{27}{22.28}{17.77}{28}
\emline{22.28}{17.77}{29}{20.55}{17.60}{30}
\emline{20.55}{17.60}{31}{18.82}{17.82}{32}
\emline{18.82}{17.82}{33}{17.19}{18.43}{34}
\emline{17.19}{18.43}{35}{15.74}{19.38}{36}
\emline{15.74}{19.38}{37}{14.53}{20.63}{38}
\emline{14.53}{20.63}{39}{13.63}{22.12}{40}
\emline{13.63}{22.12}{41}{13.09}{23.78}{42}
\emline{13.09}{23.78}{43}{12.94}{25.51}{44}
\emline{12.94}{25.51}{45}{13.17}{27.23}{46}
\emline{13.17}{27.23}{47}{13.79}{28.86}{48}
\emline{13.79}{28.86}{49}{14.75}{30.31}{50}
\emline{14.75}{30.31}{51}{16.01}{31.51}{52}
\emline{16.01}{31.51}{53}{17.51}{32.39}{54}
\emline{17.51}{32.39}{55}{20.67}{33.06}{56}
\end{picture}
\special{em:linewidth 2.4pt}
\unitlength 0.50mm
\linethickness{0.4pt}
\begin{picture}(32.67,42.56)
\emline{12.00}{10.56}{1}{12.00}{17.23}{2}
\emline{12.00}{17.36}{3}{13.00}{21.03}{4}
\emline{12.33}{40.56}{5}{12.33}{31.56}{6}
\emline{12.33}{31.56}{7}{13.19}{29.37}{8}
\emline{13.19}{29.37}{9}{14.15}{27.52}{10}
\emline{14.15}{27.52}{11}{15.21}{26.00}{12}
\emline{15.21}{26.00}{13}{16.38}{24.82}{14}
\emline{16.38}{24.82}{15}{17.66}{23.98}{16}
\emline{17.66}{23.98}{17}{19.03}{23.48}{18}
\emline{19.03}{23.48}{19}{22.10}{23.48}{20}
\emline{22.10}{23.48}{21}{23.79}{23.99}{22}
\emline{23.79}{23.99}{23}{25.58}{24.83}{24}
\emline{25.58}{24.83}{25}{28.67}{26.89}{26}
\emline{28.67}{26.89}{27}{29.92}{29.13}{28}
\emline{29.92}{29.13}{29}{30.61}{31.26}{30}
\emline{30.61}{31.26}{31}{30.73}{33.26}{32}
\emline{30.73}{33.26}{33}{30.29}{35.13}{34}
\emline{30.29}{35.13}{35}{29.00}{37.23}{36}
\emline{29.00}{37.23}{37}{27.28}{38.32}{38}
\emline{27.28}{38.32}{39}{25.60}{38.95}{40}
\emline{25.60}{38.95}{41}{23.98}{39.11}{42}
\emline{23.98}{39.11}{43}{22.42}{38.81}{44}
\emline{22.42}{38.81}{45}{20.90}{38.03}{46}
\emline{20.90}{38.03}{47}{19.44}{36.78}{48}
\emline{19.44}{36.78}{49}{18.03}{35.07}{50}
\emline{18.03}{35.07}{51}{16.67}{32.89}{52}
\end{picture}
\special{em:linewidth 0.4pt}
\unitlength 0.5mm
\linethickness{0.4pt}
\begin{picture}(47.00,40.67)
\emline{12.00}{9.67}{1}{12.00}{40.67}{2}
\emline{21.33}{32.69}{3}{23.31}{32.47}{4}
\emline{23.31}{32.47}{5}{25.18}{31.83}{6}
\emline{25.18}{31.83}{7}{26.87}{30.79}{8}
\emline{26.87}{30.79}{9}{28.30}{29.41}{10}
\emline{28.30}{29.41}{11}{29.38}{27.75}{12}
\emline{29.38}{27.75}{13}{30.08}{25.89}{14}
\emline{30.08}{25.89}{15}{30.35}{23.92}{16}
\emline{30.35}{23.92}{17}{30.19}{21.94}{18}
\emline{30.19}{21.94}{19}{29.60}{20.05}{20}
\emline{29.60}{20.05}{21}{28.61}{18.33}{22}
\emline{28.61}{18.33}{23}{27.27}{16.87}{24}
\emline{27.27}{16.87}{25}{25.64}{15.73}{26}
\emline{25.64}{15.73}{27}{23.80}{14.99}{28}
\emline{23.80}{14.99}{29}{21.84}{14.66}{30}
\emline{21.84}{14.66}{31}{19.86}{14.76}{32}
\emline{19.86}{14.76}{33}{17.95}{15.30}{34}
\emline{17.95}{15.30}{35}{16.20}{16.24}{36}
\emline{16.20}{16.24}{37}{14.70}{17.54}{38}
\emline{14.70}{17.54}{39}{13.53}{19.14}{40}
\emline{13.53}{19.14}{41}{12.73}{20.96}{42}
\emline{12.73}{20.96}{43}{12.34}{22.90}{44}
\emline{12.34}{22.90}{45}{12.39}{24.89}{46}
\emline{12.39}{24.89}{47}{12.87}{26.81}{48}
\emline{12.87}{26.81}{49}{13.77}{28.59}{50}
\emline{13.77}{28.59}{51}{15.03}{30.12}{52}
\emline{15.03}{30.12}{53}{16.59}{31.34}{54}
\emline{16.59}{31.34}{55}{18.38}{32.19}{56}
\emline{18.38}{32.19}{57}{21.33}{32.69}{58}
\emline{38.67}{31.67}{59}{40.51}{31.46}{60}
\emline{40.51}{31.46}{61}{42.26}{30.85}{62}
\emline{42.26}{30.85}{63}{43.84}{29.87}{64}
\emline{43.84}{29.87}{65}{45.15}{28.56}{66}
\emline{45.15}{28.56}{67}{46.15}{27.00}{68}
\emline{46.15}{27.00}{69}{46.78}{25.25}{70}
\emline{46.78}{25.25}{71}{47.00}{23.41}{72}
\emline{47.00}{23.41}{73}{46.81}{21.57}{74}
\emline{46.81}{21.57}{75}{46.22}{19.81}{76}
\emline{46.22}{19.81}{77}{45.25}{18.22}{78}
\emline{45.25}{18.22}{79}{43.96}{16.89}{80}
\emline{43.96}{16.89}{81}{42.40}{15.88}{82}
\emline{42.40}{15.88}{83}{40.66}{15.24}{84}
\emline{40.66}{15.24}{85}{38.82}{15.00}{86}
\emline{38.82}{15.00}{87}{36.97}{15.17}{88}
\emline{36.97}{15.17}{89}{35.21}{15.75}{90}
\emline{35.21}{15.75}{91}{33.62}{16.70}{92}
\emline{33.62}{16.70}{93}{32.28}{17.98}{94}
\emline{32.28}{17.98}{95}{31.25}{19.53}{96}
\emline{31.25}{19.53}{97}{30.59}{21.26}{98}
\emline{30.59}{21.26}{99}{30.34}{23.10}{100}
\emline{30.34}{23.10}{101}{30.49}{24.95}{102}
\emline{30.49}{24.95}{103}{31.05}{26.72}{104}
\emline{31.05}{26.72}{105}{31.99}{28.32}{106}
\emline{31.99}{28.32}{107}{33.26}{29.67}{108}
\emline{33.26}{29.67}{109}{34.79}{30.71}{110}
\emline{34.79}{30.71}{111}{36.52}{31.39}{112}
\emline{36.52}{31.39}{113}{38.67}{31.67}{114}
\end{picture}
\special{em:linewidth 2.4pt}
\unitlength 0.50mm
\linethickness{0.4pt}
\begin{picture}(56.00,42.56)
\emline{12.00}{10.56}{1}{12.00}{17.23}{2}
\emline{12.00}{17.36}{3}{13.00}{21.03}{4}
\emline{12.33}{40.56}{5}{12.33}{31.56}{6}
\emline{12.33}{31.56}{7}{13.19}{29.37}{8}
\emline{13.19}{29.37}{9}{14.15}{27.52}{10}
\emline{14.15}{27.52}{11}{15.21}{26.01}{12}
\emline{15.21}{26.01}{13}{16.38}{24.83}{14}
\emline{16.38}{24.83}{15}{17.66}{23.99}{16}
\emline{17.66}{23.99}{17}{19.03}{23.48}{18}
\emline{19.03}{23.48}{19}{22.10}{23.49}{20}
\emline{22.10}{23.49}{21}{23.79}{23.99}{22}
\emline{23.79}{23.99}{23}{25.58}{24.84}{24}
\emline{25.58}{24.84}{25}{28.67}{26.89}{26}
\emline{29.00}{37.23}{27}{27.28}{38.33}{28}
\emline{27.28}{38.33}{29}{25.60}{38.96}{30}
\emline{25.60}{38.96}{31}{23.98}{39.12}{32}
\emline{23.98}{39.12}{33}{22.42}{38.81}{34}
\emline{22.42}{38.81}{35}{20.90}{38.03}{36}
\emline{20.90}{38.03}{37}{19.44}{36.79}{38}
\emline{19.44}{36.79}{39}{18.03}{35.07}{40}
\emline{18.03}{35.07}{41}{16.67}{32.89}{42}
\emline{29.67}{36.67}{43}{32.06}{34.11}{44}
\emline{32.06}{34.11}{45}{34.00}{31.00}{46}
\emline{34.00}{31.00}{47}{35.79}{28.84}{48}
\emline{35.79}{28.84}{49}{37.59}{26.98}{50}
\emline{37.59}{26.98}{51}{39.38}{25.41}{52}
\emline{39.38}{25.41}{53}{41.18}{24.14}{54}
\emline{41.18}{24.14}{55}{43.33}{23.00}{56}
\emline{43.33}{23.00}{57}{45.80}{22.95}{58}
\emline{45.80}{22.95}{59}{47.92}{23.37}{60}
\emline{47.92}{23.37}{61}{49.69}{24.25}{62}
\emline{49.69}{24.25}{63}{52.00}{27.00}{64}
\emline{52.00}{27.00}{65}{53.33}{29.29}{66}
\emline{53.33}{29.29}{67}{54.00}{31.43}{68}
\emline{54.00}{31.43}{69}{54.01}{33.43}{70}
\emline{54.01}{33.43}{71}{52.33}{36.67}{72}
\emline{52.33}{36.67}{73}{51.15}{38.16}{74}
\emline{51.15}{38.16}{75}{46.67}{40.00}{76}
\emline{46.67}{40.00}{77}{43.93}{39.64}{78}
\emline{43.93}{39.64}{79}{41.78}{38.87}{80}
\emline{41.78}{38.87}{81}{40.21}{37.69}{82}
\emline{40.21}{37.69}{83}{39.00}{35.33}{84}
\end{picture}
\end{center}
\caption{Tadpole diagrams}\label{fig7}
\end{figure}
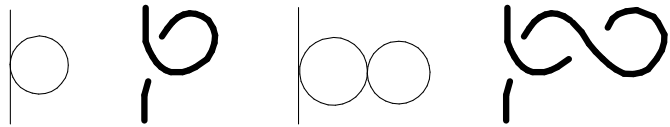

\section{Knots and M(atrix) Theory}
\setcounter{equation}{0}
In this section we discuss the limit $d\to 0$
in M(atrix) theory. Matrix models of M-theory and superstrings are obtained
by the dimensional reduction of super Yang-Mills theory in ten dimensional
spacetime to $p$ dimensions $(p=0,1,2)$ \cite {BFSS,Per,IKKT,DVV}.
The bosonic part of
the Lagrangian in the matrix approach to M-theory has the form
\begin {equation} 
                                                          \label {2.1}
S=\int (\frac{1}{2}Tr(\dot{A}_{\mu}\dot{A}_{\mu})+
\frac{g}{4N}Tr([A_{\mu}A_{\nu}][A_{\mu}A_{\nu}]))dt
\end   {equation} 
Here $A_{\mu}=A_{\mu}(t)$ are Hermitian  $N\times N$ matrices
depending on time $t$ and $\mu =1,...,d$.
One has
\begin {equation} 
                                                          \label {2.2}
\frac{1}{2}Tr([A_{\mu}A_{\nu}][A_{\mu}A_{\nu}])
= Tr(A_{\mu}A_{\nu}A_{\mu}A_{\nu})-Tr(A_{\mu}A_{\mu}A_{\nu}A_{\nu})
\end   {equation} 
The first term in (\ref {2.2}) has the form which has been discussed
in the previous section. In the limit $d\to 0$ the principal
contribution comes from the knotlike diagrams which
have one loop with Greek indices. The same reasoning   one can
apply to the IKKT matrix model \cite {IKKT} with the Lagrangian
\begin {equation} 
                                                          \label {2.3}
S=\frac{N}{2g}Tr[A_{\mu},A_{\nu}]^{2}
\end   {equation} 
If one makes the assumption on the
existence of non-zero condencate $<A_{\nu}A_{\nu}>\sim 1$ then one gets
\begin {equation} 
                                                          \label {2.4}
S_{eff}=\frac{N}{2g}(Tr(A_{\mu}A_{\mu})+Tr(
A_{\mu}A_{\nu}A_{\mu}A_{\nu})
-Tr(A_{\mu}A_{\mu}A_{\nu}A_{\nu}))
\end   {equation} 
and one can use the described diagram technique.

It is tempting to speculate that the discussion of this note
indicates that perhaps a prototypical M(atrix)  without matrix theory
$(d=0)$ exists in void. The eleven dimensional M-theory could be obtained
from this prototypical M-theory  by the decompactification of a point.
\newpage
\begin{figure}[here]
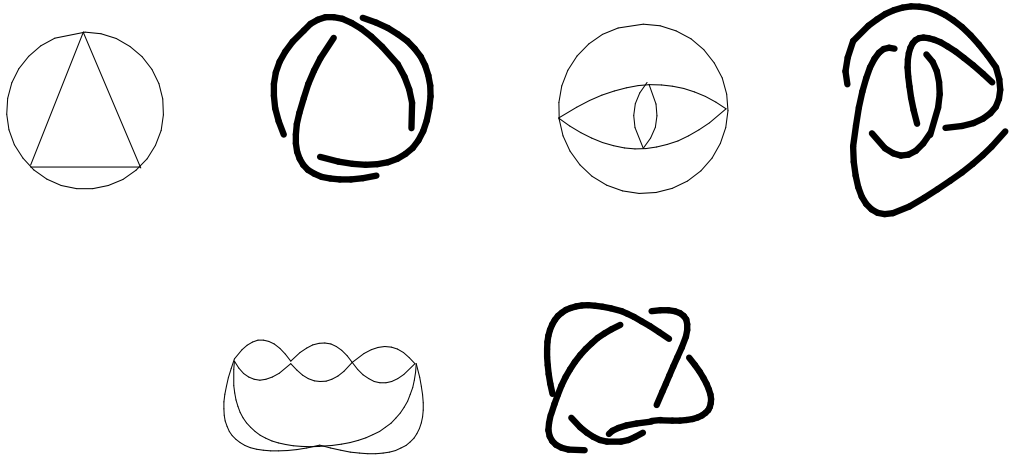

\begin{center}
\special{em:linewidth 0.4pt}
\unitlength .75mm
\linethickness{0.4pt}


\end{center}
\caption{Feynman and knot diagrams}\label{fig8}
\end{figure}

$$~$$
{\bf ACKNOWLEDGMENT}
$$~$$
The authors were stimulated by talks of V.I.Arnold and
V.A.Vassiliev on knot theory and by the paper of L.D.Faddeev and
A.J.Niemi \cite {FN}. I.A. is grateful to L.D.Faddeev for
fruitful discussions on applications of knots in field theory.
I.A. is supported
in part by  RFFI grant 96-01-00608.
I.V. is supported  in part by RFFI grant 96-01-00312.
$$~$$
\end{document}